\documentclass[aps,superscriptaddress,floatfix,nofootinbib,preprintnumbers,eqsecnum,twocolumn]{revtex4-1}
\usepackage[utf8]{inputenc}
\usepackage{lipsum, graphicx}
\usepackage{amssymb,amsmath,slashed,multirow,mathtools}
\usepackage{comment,color,placeins,amsmath,amsfonts,enumitem}
\usepackage{cancel}
\usepackage[utf8]{inputenc}

\usepackage{subfigure}

\usepackage{hyperref}
\usepackage[usenames,dvipsnames]{xcolor}
\hypersetup
{
  colorlinks,
  linkcolor={blue!80!black},
  citecolor={blue!70!black},
  urlcolor={blue!70!black}
}

\makeatletter
\renewcommand*\env@matrix[1][\arraystretch]{%
  \edef\arraystretch{#1}%
  \hskip -\arraycolsep
  \let\@ifnextchar\new@ifnextchar
  \array{*\c@MaxMatrixCols c}}
\makeatother

\def\beq{\begin{equation}}
\def\eeq{\end{equation}}

\def\bea{\begin{eqnarray}}
\def\eea{\end{eqnarray}}

\renewcommand{\slash}[1]{\not{#1}}

\begin{document}
\title{Leptophilic Axion-like Particles at Forward Detectors}
\author{Xu-Hui Jiang}
\email{jiangxh@ihep.ac.cn}
\affiliation{
Center for Future High Energy Physics, Institute of High Energy Physics, Chinese Academy of Sciences, Beijing 100049, China \\
}
\affiliation{
China Center of Advanced Science and Technology, Beijing 100190, China \\
}

\author{Chih-Ting Lu}
\email{ctlu@njnu.edu.cn}
\affiliation{
Department of Physics and Institute of Theoretical Physics, Nanjing Normal University, Nanjing, 210023, China \\ 
}

\begin{abstract} 

Leptophilic axion-like particles (ALPs) exhibit rich phenomenology, focusing exclusively on interactions between an ALP and Standard Model (SM) leptons. Through integration by parts, it is shown that both the three-point interaction, $a\bar\ell\ell$, and the four-point interaction, $a\ell^-\nu W^+$, play significant roles, making the flavor portal particularly compelling. For ALPs with masses ranging from $\mathcal O(1)$ MeV to $\mathcal O(1)$ GeV, they can contribute to exotic hadron decays. Suppressed couplings naturally extend the ALP lifetime, presenting opportunities for detection at forward detectors. In this study, we explore ALPs with both electrophilic and muonphilic scenarios. We propose an inclusive search for various hadrons that undergo exotic decays at the Large Hadron Collider (LHC). In the electrophilic scenario, long-lived ALPs are searched in the Forward Search Experiment (FASER) and its upgrading phase, FASER II. In the muonphilic scenario, where the ALP lifetime is significantly reduced due to its coupling to muons, we further investigate its detection potential at LHCb and its high-luminosity upgrade. Several benchmarks are analyzed, including electroweak-preserving, electroweak-violating and left-right softly asymmetric models, to demonstrate possible experimental constraints. 

\end{abstract}

\maketitle

\FloatBarrier

\section{Introduction}

Axion-like particles (ALPs) are typically produced through the breaking of one or more global $U(1)$ symmetries, behaving as pseudo-Nambu-Goldstone bosons~\cite{Graham:2015ouw,Choi:2020rgn}. The original new physics models involving axions were introduced to address the strong CP problem via the global $U(1)_{\text{PQ}}$ symmetry~\cite{Peccei:1977hh}, and these axions are commonly referred to as quantum chromodynamics (QCD) axions~\cite{Weinberg:1977ma,Wilczek:1977pj,Kim:1979if,Kim:2008hd}. Additionally, invisible sub-eV QCD axions are natural dark matter (DM) candidates due to the misalignment mechanism~\cite{Preskill:1982cy,Abbott:1982af,Dine:1982ah}. In contrast to QCD axions, the mass of general ALPs can range from nearly massless to the electroweak scale or higher. These ALPs may arise from the compactification of high-energy string theories~\cite{Svrcek:2006yi,Arvanitaki:2009fg,Cicoli:2012sz,Visinelli:2018utg}, models involving extra dimensions~\cite{Chang:1999si,Bastero-Gil:2021oky}, hadronization of a dark QCD~\cite{Cheng:2021kjg,Cheng:2024hvq, Cheng:2024aco}, or other new physics frameworks, which may further offer explanations for the hierarchy problem of the Higgs boson mass~\cite{Graham:2015cka}, DM or mediators to the dark sector~\cite{Arias:2012az,Zhevlakov:2022vio,Bharucha:2022lty}, cosmological issues~\cite{Jaeckel:2010ni,Jeong:2018jqe,Im:2021xoy}, and experimental anomalies~\cite{Chang:2021myh,Liu:2022tqn,Cheung:2024kml}.

In the effective field theory (EFT) framework, the interactions between ALPs and various Standard Model (SM) particles can be studied independently~\cite{Brivio:2017ije,Bauer:2017ris,Bauer:2018uxu,Ebadi:2019gij,Bauer:2020jbp,Bauer:2021mvw}. As a result, ALP search strategies depend on their mass range and coupling to SM particles. Generally, the following methods are employed to search for ALPs across different mass ranges. Light-shining-through-walls (LSW) experiments~\cite{Sikivie:1983ip,Beyer:2021mzq}, helioscopes~\cite{Irastorza:2011gs,CAST:2017uph}, and compact astrophysical objects~\cite{Kavic:2019cgk} are designed to detect ALPs in the nearly massless to MeV range. Cosmological observations can probe ALPs in the keV to GeV range, particularly those with feeble interaction strengths~\cite{Marsh:2015xka,OHare:2024nmr}. Additionally, beam dump and low-energy $e^+ e^-$ collider experiments focus on detecting ALPs in the keV to GeV range with larger interaction strengths~\cite{Riordan:1987aw,Bjorken:1988as,BaBar:2011kau,Dobrich:2015jyk,Dobrich:2019dxc,Belle-II:2020jti,Zhevlakov:2022vio,Niedziela:2020kgq,Bisht:2024hbs}. Finally, high-energy collider experiments, such as LEP, CDF, LHC, and future facilities, are optimal for detecting ALPs in the GeV to TeV range~\cite{Mimasu:2014nea,Jaeckel:2015jla,Knapen:2016moh,Brivio:2017ije,Bauer:2017ris,Bauer:2018uxu,Ebadi:2019gij,Wang:2021uyb,dEnterria:2021ljz,Zhang:2021sio,Liu:2021lan,Bao:2022onq,Han:2022mzp,Calibbi:2022izs,Cheung:2023nzg}. Notably, search results for some heavier ALPs have been reported by both the ATLAS~\cite{ATLAS:2020hii,ATLAS:2022abz,ATLAS:2023zfc,ATLAS:2023ian} and CMS~\cite{CMS:2012cve,CMS:2018erd,CMS:2021xor,CMS:2023eos,TOTEM:2023ewz,CMS:2024bnt} Collaborations. 
Despite the broad coverage of these strategies, some gaps in the ALP parameter space remain, highlighting the need for new search methods to explore these blind spots.

In this work, we focus on searching for ALP-lepton interactions using the Forward Search Experiment (FASER)~\cite{FASER:2022hcn} and Large Hadron Collider beauty (LHCb) experiment~\cite{LHCb:2008vvz}\footnote{Note that our studies differ from the LHC searches utilizing forward proton detectors for heavier ALPs~\cite{ATLAS:2023zfc,TOTEM:2023ewz}, where the two photons in the final state are detected in the central ATLAS/CMS detector. For such analyses, a dedicated Monte Carlo event generator, SuperChic~\cite{Harland-Lang:2022jwn}, can be employed.}. Typically, studies focus on the couplings between an ALP and lepton pairs, $a\ell\overline{\ell}$. However, these couplings are proportional to the charged lepton mass and are generally highly suppressed. Interestingly, as highlighted in Ref.~\cite{Altmannshofer:2022izm}, electroweak-violating scenarios give rise not only to $a\ell\overline{\ell}$ couplings but also to four-point interactions, $W$-$\ell$-$\nu$-$a$. These four-point interactions are not suppressed by the charged lepton mass and, more importantly, processes involving these interactions exhibit unique energy enhancements~\cite{Altmannshofer:2022izm,Lu:2022zbe,Lu:2023ryd,Buonocore:2023kna,Wang:2024zky}. For instance, the three-body decay rates of charged mesons into an ALP, a charged lepton, and a neutrino in the electroweak-violating scenario are much larger than those in the electroweak-preserving scenario. Thus, we can explore the structure of electroweak symmetry in ALP-lepton interactions, even within the EFT framework.

For a light ALP with feeble interactions to charged leptons, it naturally behaves as a long-lived particle (LLP). FASER (FASER II) is a cylinder shape detector with the length $L = 1.5$ m ($L = 5$ m) and radius $R = 0.1$ m ($R = 1$ m) which is $480$ m downstream from the ATLAS interaction point. The integrated luminosity for FASER and FASER II will be $300$ fb$^{-1}$ and $3000$ fb$^{-1}$, respectively. Therefore, the FASER detector is specifically designed to search for such LLPs. For example, studies in Refs.~\cite{FASER:2023tle,FASER:2024bbl} have explored light dark photons and ALPs decaying into final states involving electrons, positrons and photons. The LHCb detector is composed of a forward spectrometer and planar detectors, and its size is 21 m long, 13 m wide, and 10 m high. The integrated luminosity for LHCb is around several inverse femtobarns, but has been expected to be 300 fb$^{-1}$. Similarly, the LHCb detector is well-suited for searching for light ALPs, thanks to its precise vertex reconstruction capabilities and flexible trigger system. Consequently, various search strategies at LHCb have been investigated in Refs.~\cite{LHCb:2014osd,LHCb:2015nkv,LHCb:2016awg,LHCb:2017trq,LHCb:2019vmc,Borsato:2021aum,Gorkavenko:2023nbk}. In this work, we focus on the scenarios of electrophilic and muonphilic ALPs and study exotic meson decays with both two-body and three-body final states involving the ALP, which subsequently decays into $e^+e^-$, $\mu^+\mu^-$, or $\gamma\gamma$.

The remainder of this paper is organized as follows. In Sec.~\ref{sec:alpproduction}, we review the ALP-lepton interactions, with a particular focus on the ALP decay and production modes. In Sec.~\ref{sec:es}, we outline the search strategies and present the analyses for electrophilic ALPs using the FASER detector and its upgraded version, FASER II. In Sec.~\ref{sec:mus}, we extend our studies to muonphilic ALPs, analyzing their prospects at the FASER, FASER II, and LHCb detectors. Finally, we summarize our findings in Sec.~\ref{sec:con}. Additional details regarding the derivations of ALP-lepton interactions are provided in Appendix~\ref{app: interaction}.

\section{ALP-Lepton Interactions: Productions and Decays}\label{sec:alpproduction}

The ALP, denoted as $a$, arises from the breaking of the global $U(1)_{\text{PQ}}$ symmetry~\cite{Peccei:1977hh}, which introduces a shift symmetry in the Lagrangian, $a(x)\to a(x) +\text{const}$. Utilizing this property, we can express the ALP-lepton interactions as:  
\begin{equation}
\label{eq:Lag}
\partial_{\mu} a J^{\mu}_{\text{PQ},\ell} ~.
\end{equation}
Here, $J^{\mu}_{\text{PQ},\ell}$ is the general lepton current associated with the global $U(1)_{\text{PQ}}$ symmetry, given by~\cite{Altmannshofer:2022izm,Bertuzzo:2022fcm,Lu:2022zbe}, 
\begin{equation} 
\label{eq:Jint}
J^{\mu}_{\text{PQ},\ell} = \frac{c^V_{\ell}}{2\Lambda}\overline{\ell}\gamma^{\mu}\ell + \frac{c^A_{\ell}}{2\Lambda}\overline{\ell}\gamma^{\mu}\gamma_5\ell + \frac{c_{\nu}}{2\Lambda}\overline{\nu_{\ell}}\gamma^{\mu} P_L \nu_{\ell} \,, 
\end{equation} 
where $\Lambda$ represents the $U(1)_{\text{PQ}}$ symmetry-breaking scale, and $c^V_{\ell}$, $c^A_{\ell}$, and $c_{\nu}$ are dimensionless couplings. The symbols $\ell$ and $\nu$ denote charged leptons and neutrinos, respectively. After integration by parts and using the equations of motion to transform $\partial_{\mu} a J^{\mu}_{\text{PQ},\ell}$ into $a\partial_{\mu}J^{\mu}_{\text{PQ},\ell}$, the full Lagrangian for ALP-lepton interactions is written as~\cite{Altmannshofer:2022izm,Lu:2022zbe}:  
\begin{align} 
& a ~\partial_{\mu}J^{\mu}_{\text{PQ},\ell} = i c^A_{\ell}\frac{m_{\ell}}{\Lambda}~a\overline{\ell}\gamma_5\ell \label{eq:int} \\
& + \frac{\alpha_{\text{em}}}{4\pi\Lambda} \bigg[  \frac{ c^V_{\ell} -c^A_{\ell} + c_{\nu}}{4 s^2 _W}~a W^{+}_{\mu\nu}\tilde W ^{-,\mu\nu} \notag \\ 
& + \frac{c^V_{\ell} - c^A_{\ell} (1 -4 s^2_W)}{2s _W c_W}~a F_{\mu\nu}\tilde{Z} ^{\mu\nu} - c^A_{\ell}~a F_{\mu\nu} \tilde{F}^{\mu\nu} + \notag \\ 
& \frac{c^V_{\ell} (1 -4 s^2_W) -c^A_{\ell} (1 -4 s^2_W +8 s^4_W)  + c_{\nu}}{8 s^2_W c^2_W}~a Z_{\mu\nu}\tilde{Z}^{\mu\nu}\bigg]  \notag \\ 
& - \frac{ig}{2\sqrt{2}\Lambda}(c^A_{\ell} - c^V_{\ell} + c_{\nu})~a (\bar\ell \gamma^{\mu} P _L \nu) W_{\mu}^{-} ~+~\text{h.c.}  \,. \notag 
\end{align}
Here, $m_{\ell}$ is the charged lepton mass, $\alpha_{\text{em}}$ is the fine structure constraint, $s_W$ and $c_W$ are the sine and cosine of the weak mixing angle, respectively, and $g$ is the weak coupling constant. The symbols $W^{\pm}_{\mu\nu}$, $F_{\mu\nu}$, and $Z_{\mu\nu}$ are the field strength tensors for $W^{\pm}$, $\gamma$, and $Z$, respectively. The dual field strength tensor is defined as $\tilde{F}_{\mu\nu}=\frac{1}{2}\epsilon_{\mu\nu\rho\sigma}F^{\rho\sigma}$. Details of the derivation from Eq.~(\ref{eq:Lag}) to Eq.~(\ref{eq:int}) are provided in Appendix~\ref{app: interaction}~\footnote{The four-point interaction term differs from that in Ref.~\cite{Altmannshofer:2022izm} by an extra minus sign. This arises due to different conventions of the covariant derivative being used in calculations.}. 

According to the ALP-lepton interactions in Eq.~(\ref{eq:int}), the ALP can decay directly to a pair of leptons or a pair of gauge bosons due to the chiral anomaly and one-loop triangle Feynman diagrams. For ALPs with masses well below the electroweak scale, the two leading partial decay widths can be expressed as~\cite{Bauer:2017ris,Bauer:2018uxu,Chang:2021myh}
\begin{align}
&\label{eq:Gammall} \Gamma_{a\rightarrow\ell^{+}\ell^{-}} = \frac{(c^A_{\ell})^2 m^2_{\ell} m_a}{8\pi\Lambda^2}\sqrt{1-\frac{4m^2_{\ell}}{m^2_a}} \,, \\ 
&\label{eq:Gammaaa} \Gamma_{a\rightarrow\gamma\gamma} = \frac{m^3_a}{64\pi}\left(\frac{\alpha_{\text{em}}}{\pi}\frac{c^A_{\ell}}{\Lambda}\lvert 1 - {\cal F} (\frac{m^2_a}{4m^2_{\ell}})\rvert \right)^2  \,,
\end{align}
where the loop functions are given by ${\cal F} (z > 1) = \frac{1}{z}\text{arctan}^2\left(\frac{1}{\sqrt{1/z -1}}\right)$ and ${\cal F} (z < 1) = \frac{1}{z}\text{arcsin}^2\left(\sqrt{z}\right)$, respectively. Both dilepton and diphoton are important decay channels to be searched for at prompt and far detectors. In the following sections, we will focus on the search for light ALPs using forward detectors, including FASER, FASER II, and LHCb.

Due to the presence of $aVV^{(\prime)}$ terms arising from the chiral anomaly and the novel four-point, $aW\ell\nu$, terms which depend on the choices of the coefficients $c^V_{\ell}$, $c^A_{\ell}$, and $c_{\nu}$, we identify two specific scenarios related to the electroweak symmetry structure~\cite{Altmannshofer:2022izm}: 
\begin{align} 
& \text{Electroweak Violating}~ (\text{EWV}) : c^V_{\ell} = c_{\nu} = 0, c^A_{\ell} \neq 0, \notag \\ 
& \text{Electroweak Preserving}~ (\text{EWP}) : c_{\nu} = 0, c^V_{\ell} = c^A_{\ell} \neq 0\,. 
\label{eq:EWV_EWP}
\end{align} 
In the EWV scenario, the lepton current in Eq.~(\ref{eq:Jint}) is purely axial-vector, while in the EWP scenario, it corresponds to right-handed coupling. Additionally, in the EWV scenario, all terms in Eq.~(\ref{eq:int}) are involved. In contrast, for the EWP scenario, only the $a\ell\ell$, $a\gamma\gamma$, $aZZ$, and $aZ\gamma$ interactions in Eq.~(\ref{eq:int}) remain. 

In addition to the EWV and EWP scenarios, we further explore another intriguing case, known as the left-right-softly-asymmetric (LRSA) scenario. A left-right symmetric UV completion is quite general, but to allow the ALP to decay into a pair of leptons, the left-right symmetry must be broken to some extent. To quantify this breaking, we define the following variable:  
\begin{equation}
    \Delta \equiv \bigg | \frac{c_L-c_R}{c_L+c_R}\bigg |~,
\end{equation}
where $c_L$ ($c_R$) represents the coupling of the ALP to the left-chiral (right-chiral) leptonic current. Alternatively, $\Delta$ can be equivalently parameterized as $|\frac{c_\ell^A}{c_\ell^V}|$. A nearly non-chiral UV completion corresponds to $\Delta \ll 1$. In this scenario, the branching ratios for ALP production are naturally enhanced, as indicated in Eqs.~(\ref{eq:gdd}) and~(\ref{eq:3body}) below. Meanwhile, the ALP decay rate keeps almost unchanged. At tree-level, the decay of an ALP into either diphoton or dilepton final states depends purely on $c_\mu^A$, as shown by Eq.~(\ref{eq:int}). The role of $c_\mu^V$ becomes relevant only when electroweak gauge bosons are involved. However, such contributions are heavily suppressed by the loop factor $(\frac{m_a}{m_W})^{2n}$, where $n$ depends on the specific loop structures. This scenario is particular attractive because the non-vanishing $c_\mu^V$ can enhance event rates without shortening the ALP lifetime. Consequently, this provides a novel opportunity to observe signals at either far or prompt detectors. In this study, we adopt $\Delta = 0.1$ as a benchmark, which corresponds to $c_\mu^A = -0.1 c_\mu^V$ and $c_\nu =0$.

Based on the above discussions, we further classify whether the production of ALPs from exotic decays of muons and mesons are sensitive to the four-point $aW\ell\nu$ interaction term. In the case of four-body muon decays, $\mu^+ \to e^+ \nu_e \bar\nu_\mu a$, and three-body charged meson decays, $P^+ \to \ell^+ \nu a$, with $P = \pi$, $K$, $D$, and $D_s$, novel energy enhancements from the four-point $aW\ell\nu$ interaction term were first identified in Ref.~\cite{Altmannshofer:2022izm}. Consequently, we expect these exotic decay rates to be significantly larger in the EWV scenario than in the EWP scenario. On the other hand, in the case of two-body flavor-changing neutral current (FCNC) meson decays, such as $B\to K^{(\ast)} a$, $K^{\pm}\to\pi^{\pm} a$, and $K_{L,S}\to\pi^0 a$, the absence of the $aWW$ interaction in the EWP scenario suggests that these two-body FCNC meson exotic decay rates would be suppressed compared to those in the EWV scenario. 

\begin{table}[ht!]
\begin{center}\begin{tabular}{|c|c|c|}
\hline 
\textbf{Scenario} & \textbf{Electrophilic} & \textbf{Muonphilic} \\
\hline
\hline
\multirow{2}*{\textbf{4-body}} & $\mu^{+}\to e^{+}\nu_e\bar\nu_\mu a$ & $\mu^{+}\to e^{+}\nu_e\bar\nu_\mu a$ \\
                               ~ & $a\to \gamma\gamma,~ e^+ e^-$ & $a\to \gamma\gamma$  \\
\hline 
\multirow{2}*{\textbf{3-body}} & $P^{+}\to e^{+}\nu_e a$ & $P^{+}\to\mu^{+}\nu_{\mu}a$  \\
                               ~ & $a\to \gamma\gamma, ~e^+ e^-$ & $a\to\gamma\gamma, ~\mu^+\mu^-$  \\
\hline 
\multirow{2}*{\textbf{2-body}} & $M_1\to M_2 a$ & $M_1\to M_2 a$  \\
                               ~ & $a\to \gamma\gamma,~ e^+ e^-$ & $a\to \gamma\gamma, ~\mu^+\mu^-$ \\
\hline 
\end{tabular} \caption{The relevant productions and decays of ALPs for electrophilic and muonphilic ALPs. The notations are defined as $P = \pi, K, D, D_s$, and $\ell = e, \mu$, as well as $(M_1, M_2) = (B, K^{\ast}), (K^{+}, \pi^{+}), (K_{L,S},\pi^0)$. Notably, the ALP is necessary to be lighter than $(m_P- m_\ell )$ and $(m_{M_1}-m_{M_2})$ in 3-body and 2-body decays, respectively. Moreover, $a\to \ell^+\ell^-$ channel is turned on only if $m_a > 2m_\ell$.}
\label{tab:pro_dec}
\end{center}
\end{table}

In this paper, we follow Ref.~\cite{Altmannshofer:2022izm} and only explore the three-body and two-body meson decays, for simplicity\footnote{Four-body muon decays can produce ALPs with masses less than approximately $100$ MeV, which are already subject to strong constraints from experiments such as the E137 beam-dump experiment~\cite{Bjorken:1988as} and observations of supernova SN1987A~\cite{Carenza:2021pcm}.}. Basically, there are two primary scenarios\footnote{Tau-philic ALPs are unlikely to be produced due to the significantly heavier $\tau$-lepton compared to the two lighter flavors. Therefore, they fall beyond the scope of this study and will not be further investigated.}: electrophilic and muonphilic ALPs, both of which can be searched for using far detectors. Both scenarios can be realized by fruitful ultraviolet (UV) completion models, as discussed in the Appendix of Ref.~\cite{Altmannshofer:2022izm}. Here, the relevant productions and decays of ALPs are summarized in Table~\ref{tab:pro_dec}.

It is important to highlight that the approach of this work differs from that in Ref.~\cite{Altmannshofer:2022izm}. Instead of establishing bounds for ALP-lepton interactions based on precision measurements of meson decays, we focus on directly searching for ALPs with the forward detectors, similar to the approach in Refs.~\cite{Buonocore:2023kna,Wang:2024zky}. These two different methodologies are complementary to one another. 

As summarized in Table~\ref{tab:pro_dec}, various hadron decays contribute to our analysis. The production of ALPs has been extensively studied in Ref.~\cite{Altmannshofer:2022izm}. Building on this foundation, we present all relevant branching ratios for exotic meson decays used in this work. We begin with the 2-body decays as follows: 
\begin{align}
\text{BR}(B^0 \to K^\ast a) =& \frac{|g_{bs}|^2}{64\pi} \frac{m_B^3}{\Lambda^2} |A_0^{B\to K^\ast}(m_a^2)|^2\times \notag \\
& \lambda^{\frac{3}{2}}\bigg (\frac{m_{K^\ast}^2}{m_B^2},~\frac{m_a^2}{m_B^2}\bigg )~,
\end{align}
\begin{align}
\text{BR}(K^+ \to \pi^+a) =& \frac{|g_{sd}|^2}{64\pi} \frac{m_{K^+}^3}{\Lambda^2}\bigg (1-\frac{m_{\pi^+}^2}{m_{K^+}^2}\bigg )^2\times \notag \\ 
& \lambda^{\frac{1}{2}}\bigg (\frac{m_{\pi^+}^2}{m_{K^+}^2},~\frac{m_a^2}{m_{K^+}^2}\bigg )~, 
\end{align}
\begin{align}
\text{BR}(K_L \to \pi^0a) =& \frac{|\text{Im}(g_{sd})|^2}{64\pi} \frac{m_{K^0}^3}{\Lambda^2}\bigg (1-\frac{m_{\pi^0}^2}{m_{K^0}^2}\bigg )^2\times \notag \\ 
& \lambda^{\frac{1}{2}}\bigg (\frac{m_{\pi^0}^2}{m_{K^0}^2},~\frac{m_a^2}{m_{K^0}^2}\bigg )~, \\
\text{BR}(K_S \to \pi^0a) =& \frac{\tau_{K_S}}{\tau_{K_L}}\text{BR}(K_L \to \pi^0a)~,
\end{align}
where $\lambda (a,~b)=1+a^2+b^2-2(a+b+ab)$, and $A_0^{B\to K^\ast}$ is the form factor which can be found in Refs.~\cite{Horgan:2013hoa, Horgan:2015vla}. Furthermore, $g_{d_id_j}$ could be formulated as:
\begin{align}
g_{d_id_j} =& -\frac{g^2}{16\pi^2} V_{ti}^\ast V_{tj} \bigg [ \frac{g^2}{16\pi^2}\frac{3}{8}(c_\ell^A-c_\ell^V-c_\nu) F(x_t) \notag \\
& + \frac{g^{\prime 4}}{(16\pi^2)^2}\frac{17}{96}(c_\ell^A+c_\ell^V)x_t \log^2 (\frac{\Lambda^2}{m_t^2})\bigg]~, \label{eq:gdd}
\end{align}
with $g$ and $g^\prime$ denoting electroweak couplings. $x_t$ is defined as $x_t = \frac{m_t^2}{m_W^2}$, where $m_t$ and $m_W$ are masses of top quark and $W^\pm$ boson, respectively. The function $F(x)$ parameterizes the loop effect, which takes the following form:
\begin{equation}
    F(x)=x \bigg(\frac{1}{1-x} + \frac{x \log x}{(1-x)^2} \bigg )~.
\end{equation}
Particularly, the first term in Eq.~(\ref{eq:gdd}) corresponds to a two-loop contribution, where a top loop and an anomaly triangle are involved. In addition, the corrections from higher loop effects are shown as the second term. Notably, we regard the decay of $K_S$ differs from $K_L$ only in its lifetime, which is consistent with the treatment in other works~\cite{Buonocore:2023kna, Berger:2024xqk}.

\begin{figure*}[ht!]
\includegraphics[width=0.8\linewidth]{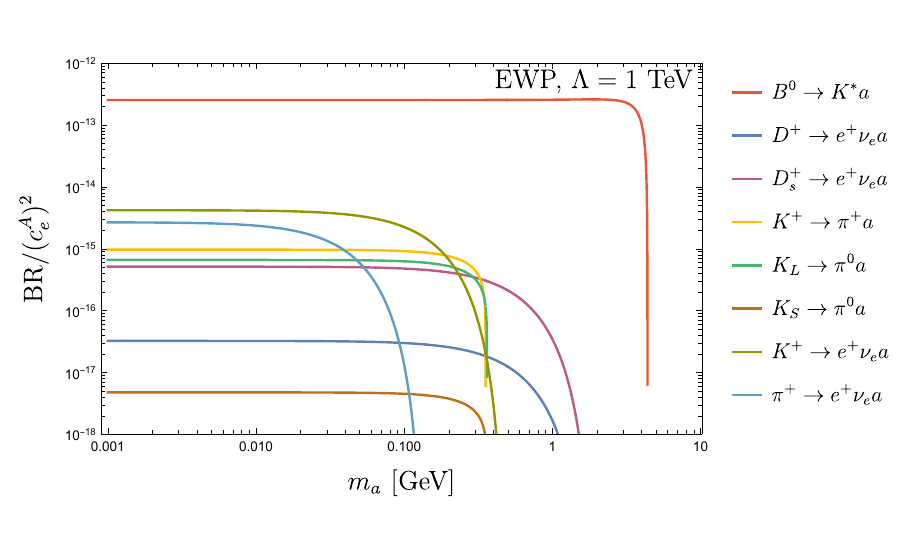}\qquad 
\includegraphics[width=0.8\linewidth]{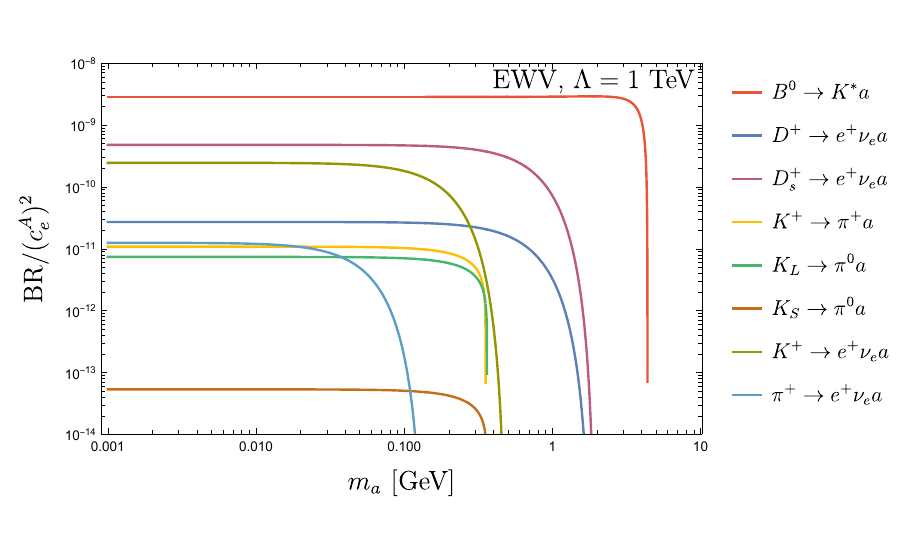} 
\caption{The branching ratios of ALP producitons via various exotic meson decay processes. \textbf{Upper panel:} The EWP scenario. \textbf{Lower panel:} The EWV scenario.}
\label{fig:electrobraxiongeneration}
\end{figure*} 

The 3-body decays are considered next. When the parent particle is significantly heavier than the lepton, we have the following relation: 
\begin{align}
& \frac{\text{BR}(P^+ \to \ell^+ \nu_\ell a)}{\text{BR}(P^+ \to \mu^+ \nu_\mu)} = \frac{m_P^4}{1536\pi^2 m_\mu^2\Lambda^2}\bigg (1- \frac{m_\mu^2}{m_P^2}\bigg )^{-2}\times \notag \\
& \bigg [ (c_\ell^A-c_\ell^V+c_\nu)^2 f_0(x_P) + \frac{16 m_\ell^2}{m_P^2}c_\ell^{A2}f_1(x_P)\bigg]~, \label{eq:3body}
\end{align}
with $x_P=\frac{m_a^2}{m_P^2}$ and $f$s in the forms:
\begin{align}
    f_0(x) &= 1-8x+8x^3-x^4 - 12x^2\log x~,\\
    f_1(x) &= 1+ 9x -9x^2 -x^3 + 6x(1+x) \log x~.
\end{align} 
In the EWV and LRSA scenarios, higher loop effects are sub-leading. Consequently, the branching ratios discussed above are insensitive to the specific lepton flavors. Assuming lepton flavor universality (LFU) between the first two generations of leptons, an electrophilic ALP has approximately the same production rates as a muonphilic one. However, this differs in the EWP scenario. Specifically, the two-loop contributions vanish in the EWP scenario, leading to a more suppressed rate of exotic meson decays. For two-body exotic meson decays, the branching ratios remain insensitive to lepton flavors, while three-body decays depend explicitly on $m_\ell^2$, as shown in Eq.~(\ref{eq:3body}). As a result, muonphilic ALP production rates can be obtained by rescaling those of electrophilic ALPs by a factor of $(\frac{m_{\mu}}{m_e})^2$. As an example, we consider the electrophilic scenario. The PDG data~\cite{ParticleDataGroup:2024cfk} are used as inputs. We present the branching ratios for various meson decays in Fig.\ref{fig:electrobraxiongeneration}.

Furthermore, we emphasize that the above formulae remain valid even when the ALP mass falls below $2m_\ell$. In this scenario, the charged leptons act as internal loop lines to facilitate ALP production, while the corresponding dilepton decay is switched off, leading to potentially distinct signatures at the detectors. Additional details will be discussed in the following sections.

\section{Electrophilic ALPs}\label{sec:es}

\begin{figure*}[ht!]
\includegraphics[width=0.45\linewidth]{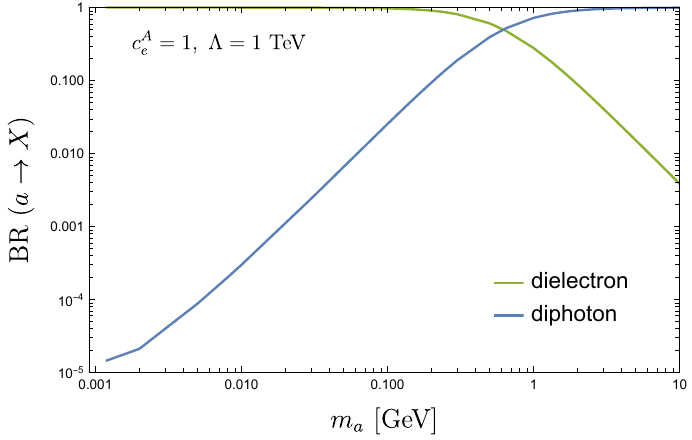}\qquad 
\includegraphics[width=0.45\linewidth]{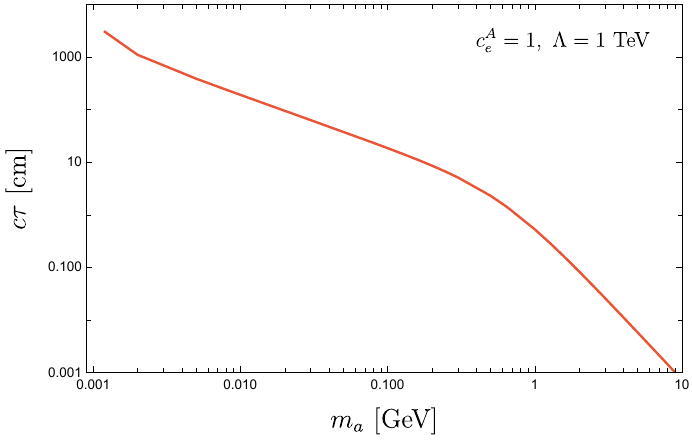} 
\caption{\textbf{Left panel:} The branching ratios of an electrophilic ALP decaying to dielectron and diphoton, where $c_e^A = 1$ and $\Lambda = 1$ TeV are applied for illustration. \textbf{Right panel:} The proper lifetime of the ALP varies with the ALP mass.}
\label{fig:electrobr}
\end{figure*}

An electrophilic ALP can play a significant role in exotic decays of various hadrons. The branching ratios of these hadrons may deviate substantially from SM predictions, indicating potential signals of beyond the SM. In this study, we consider an electrophilic ALP with a mass ranging from twice the electron mass, $2m_e$, to the mass difference $m_{B^0}-m_{K^\ast}$, including the process $B^0 \to K^\ast a$. For simplicity, the cut-off scale in Eq.~(\ref{eq:Jint}) is set to $\Lambda = 1$~TeV as a benchmark point. Based on Eqs.~(\ref{eq:Gammall}) and~(\ref{eq:Gammaaa}), the decays $a\to e^+e^-$ and $a\to \gamma\gamma$ are kinematically allowed. In Fig.~\ref{fig:electrobr}, we show the branching ratios and proper lifetime of the ALP as functions of its mass assuming the coupling constant $c_e^A = 1$ for easy scaling in subsequent discussions. For lower ALP masses, the dielectron channel dominates, while the diphoton channel takes precedence at higher masses. The transition between these two regimes occurs around $0.6$ GeV, where the ALP's proper lifetime is approximately $10 / (c_e^A)^2$ cm. Importantly, a coupling constant $c_e^A$ of $\mathcal O(1)$ can extend the ALP lifetime to macroscopic distances, making it a promising candidate for an LLP.

\begin{figure*}[ht!]
\includegraphics[width=0.45\linewidth]{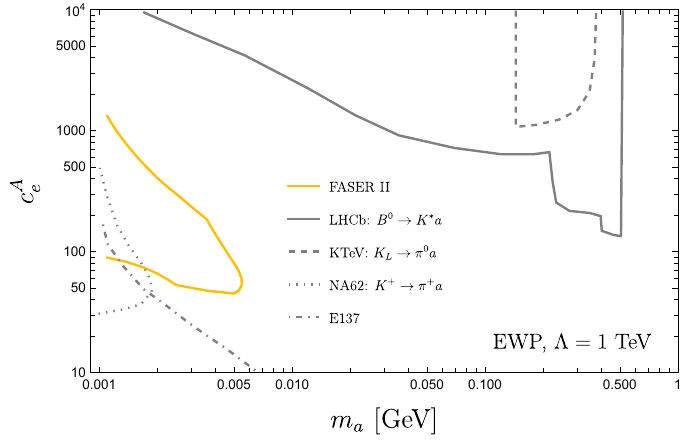}\qquad 
\includegraphics[width=0.45\linewidth]{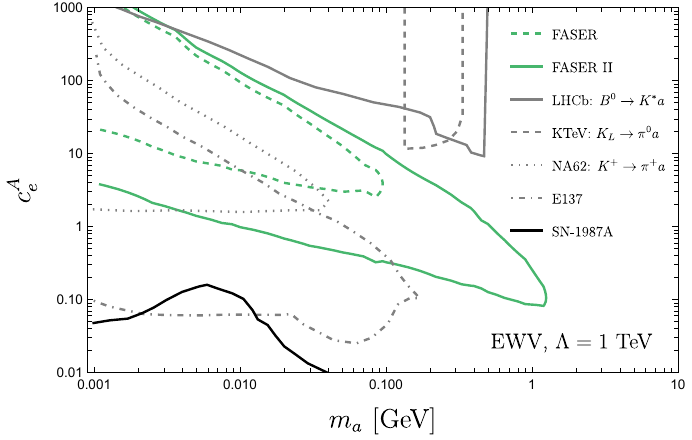}\qquad 
\includegraphics[width=0.45\linewidth]{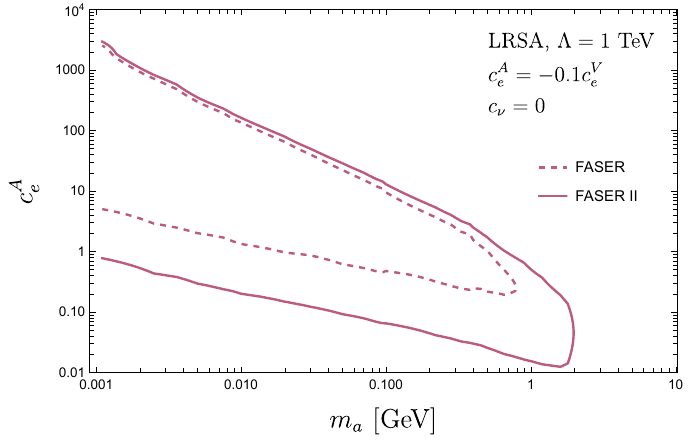} 
\caption{The $95\%$ confidence level projections that FASER (dashed line) and FASER II (solid lines) may reach, with an inclusive search for electrophilic ALPs. The EWP, EWV, and LRSA scenarios are shown in the left, right and lower panels, respectively. Projections from other experiments are also provided for comparison, including FCNC decays at prompt detectors~\cite{KTeV:2003sls, LHCb:2015ycz, NA62:2020xlg}, the electron beam dump experiment E137~\cite{Bjorken:1988as}, and constraints derived from supernova SN-1987A~\cite{Carenza:2021pcm}.}
\label{fig:electrofaser}
\end{figure*}
 
Such signals from ALP decays would be detected by far detectors, such as FASER. We inclusively consider the exotic decays of various hadrons, as summarized in Table.~\ref{tab:pro_dec}. The hadron productions and decays are simulated using Pythia 8~\cite{Sjostrand:2014zea} for heavy mesons ($B^0$, $D^\pm$, and $D_s^\pm$) and EPOS LHC~\cite{Pierog:2013ria} for light mesons ($\pi^\pm$, $K^\pm$ and $K_{L,S}$)\footnote{Currently, only Pythia 8 supports simulations of
heavy mesons, and for light mesons, as demonstrated in the Appendix of Ref.~\cite{Foroughi-Abari:2020qar}, the EPOS LHC simulation exhibits excellent agreement with LHC measurements.}. The branching ratios of ALP decays are referenced from those presented in Sec.~\ref{sec:alpproduction}. All physical quantities are taken from their PDG values~\cite{ParticleDataGroup:2024cfk}. The FASER environment is simulated using the foresee package~\cite{Kling:2021fwx}. To ensure that the background-free assumption remains valid, we follow the first FASER measurement~\cite{FASER:2023tle} and require the ALP decay vertex within the detector and the total energy inside FASER detectors to exceed $500$ GeV. Here both dielectron and diphoton final states are searched for in our analysis. For the background-free analysis, we adopt the $N=3$ hypothesis for the $95\%$ confidence level~\footnote{Under the background-free hypothesis, the signal events are expected to follow a Poisson distribution. A $95\%$ confidence level corresponds to the condition that at least one event is observed when the expected average is three.}. The limits from FASER and its upgraded phase, FASER II, are projected as dashed and solid curves, respectively, in Fig.~\ref{fig:electrofaser}, with the orange, green and violet curves representing the EWP, EWV and LRSA scenarios, respectively. For comparison, Fig.~\ref{fig:electrofaser} also includes projections from other experiments searching for light leptophilic ALPs. These include FCNC decays at prompt detectors~\cite{KTeV:2003sls, LHCb:2015ycz, NA62:2020xlg}, the electron beam dump experiment E137~\cite{Bjorken:1988as}, and constraints from supernova SN-1987A~\cite{Carenza:2021pcm}.

Due to the limited production rates of ALPs, FASER is unlikely to capture any relevant signals in the EWP scenario. In contrast, the increased luminosity and larger decay vessel at FASER II make it possible to constrain the model parameters in the EWP scenario, albeit within a narrow range where the ALP mass is lighter that $5$ MeV, but the coupling $c_e^A \sim \mathcal O(10^2)$. Remarkably, as shown in Fig~\ref{fig:electrofaser}, a slight overlap is observed between the search regions of $K^+$ decays and the E137 beam dump experiment, whereas neither $K_L$ nor $B^0$ FCNC decays at prompt detectors have reached the region accessible to FASER II.

The EWV scenario is much more optimistic. In FASER experiment, the $95\%$ confidence level limit covers the region where $c_e^A$ ranges from $10$ to $1000$, and the mass can reach up to $0.1$ GeV. Specifically, the dielectron channel dominates, and thus diphoton signals are not expected to be significant for the time being. However, this situation will change significantly when moving to the FASER II phase. As shown in Fig.~\ref{fig:electrofaser}, a better projection is obtained. The coupling $c_e^A$ can be constrained to $\mathcal O(1)$, with the ALP mass increasing up to $1$ GeV. To achieve this limit, both dielectron and diphoton signals play crucial roles in FASER II experiment, as indicated in Fig.~\ref{fig:electrobr}. Furthermore, when the ALP mass exceeds $1$ GeV, even FASER II loses its sensitivity. This can be explained as follows: as shown in the right panel of Fig.~\ref{fig:electrobr}, when the ALP mass exceeds $1$ GeV, its lifetime decreases significantly, causing it to decay promptly and making it difficult to detect in far detectors like FASER and its upgrade phase, FASER II.

In the LRSA scenario, ALP production rates are enhanced due to the non-vanishing $c_e^V$, which leads to an expanded parameter space for $m_a$ and $c_e^A$, resulting in enlarged contours at the $95\%$ confidence level. In the FASER experiment, unlike the EWP and EWV scenarios, the diphoton decay channel of ALPs becomes a relevant and promising signal to target within the detector. FASER II extends the search to regions with smaller couplings and slightly larger ALP masses, reaching approximately $2$ GeV. However, the search for heavier ALPs remains challenging, primarily due to their shortened lifetimes caused by the diphoton decay channel.

Our results are largely consistent with those in Ref.~\cite{Buonocore:2023kna}, though some minor differences in contour shapes are observed. Several factors contribute to these differences. First, this study does not account for higher-order effects in event generation and showering. Additionally, the event selection criteria differ from those in Ref.~\cite{Buonocore:2023kna}. Specifically, following the search strategies employed by the FASER collaboration~\cite{FASER:2023tle}, we apply a stricter cut, requiring the total energy inside FASER and FASER II to exceed $500$ GeV.

Furthermore, as illustrated in Fig.~\ref{fig:electrofaser}, FASER demonstrates unique potential to explore a previously uncovered region between the sensitivity projections of $K^+$ decays at NA62 and $B^0$ decays at LHCb. This reach will be significantly extended at FASER II. For regions favoring a small ALP mass (lighter than $0.1$ GeV) and small coupling (weaker than $\mathcal{O}(10)$), FASER II is expected to nearly cover the projection of NA62 and partially overlap with the beam dump experiment E137. However, FASER II loses sensitivity as the coupling becomes larger (shortening the lifetime), while this regime is already probed by $B^0$ or $K_L$ decays at prompt detectors. 
Additionally, we compare the sensitivity of FASER and FASER II to other proposals, such as SHiNESS~\cite{Wang:2024zky}. The two-year reach of SHiNESS is entirely independent of FASER, and it will intersect with FASER II only in limited cases where $m_a$ and $c_e^A$ reach approximately $\mathcal{O}(10)$ MeV and $\mathcal{O}(1)$, respectively. Notably, the region where $m_a > 10$ MeV and $c_e^A < 10$ remains largely unexplored. Encouragingly, FASER II is expected to probe this blind spot in the near future. 

\section{Muonphilic ALPs}\label{sec:mus}

\begin{figure*}[ht!]
\includegraphics[width=0.45\linewidth]{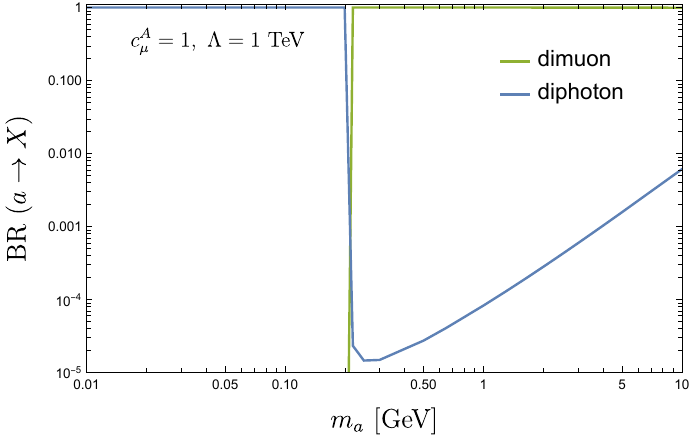}\qquad 
\includegraphics[width=0.45\linewidth]{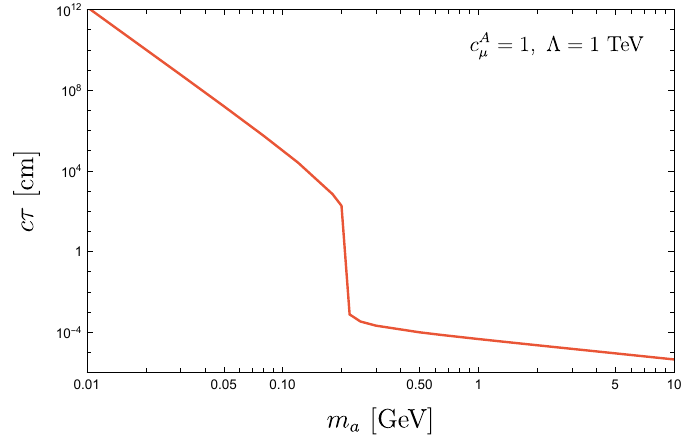} 
\caption{\textbf{Left panel:} The branching ratios of a muonphilic ALP decaying to dimuon and diphoton, where $c_\mu^A = 1$ and $\Lambda = 1$ TeV are fixed for illustration. \textbf{Right panel:} The proper lifetime of the ALP varies with the ALP mass. }
\label{fig:muonbr}
\end{figure*}

Theoretically, a muonphilic ALP is quite similar to its electrophilic counterpart. However, the threshold mass $2m_\mu$ is much bigger than that of the electrophilic ALP. To explore a richer phenomenology, we decide to extend the mass range, starting from $0.05$ GeV. The decays $a \to \mu^+\mu^-$ and $a \to\gamma\gamma$ are both studied. The branching ratios and proper lifetime of the ALP are presented in Fig.~\ref{fig:muonbr}, with the coupling constant $c_\mu^A$ set to unity.

The decay topology of a muonphilic ALP is quite simple. Above the threshold $2m_\mu$, the dimuon channel dominates due to the significantly heavier mass of the muon compared to the electron, while the diphoton channel contributes less than $10^{-3}$ and can reasonably be neglected at this stage. However, when the ALP mass drops below the threshold, the dimuon channel is forbidden, and the ALP can only decay into a pair of photons. Notably, the diphoton decay rate differs from the dimuon decay rate. As shown in the right panel of Fig.~\ref{fig:muonbr}, the ALP proper lifetime experiences a sharp drop near the threshold $m_a\sim 2m_\mu$. Generally speaking, above the threshold, assuming LFU holds between electron and muon, the lifetime of a muonphilic ALP is roughly four orders of magnitude shorter than that of an electrophilic ALP, with the suppression factor proportional to $(\frac{m_e}{m_\mu})^2$. This makes a muonphilic ALP more likely to decay before reaching the FASER detector. Opportunities may arise at prompt detectors like LHCb, if notably, the ALP lifetime reaches $1$ cm, which can occur when the coupling $c_\mu^A \sim \mathcal O(10)$ and the ALP mass is around $0.5$ GeV. Below the threshold, the diphoton decay ensures that the ALP remains sufficiently long-lived. The corresponding signals can be searched for at FASER, complementing the searches for heavier-mass ALPs at prompt detectors. In summary, in scenarios such as EWP and EWV, we observe a competition between the ALP production rate and its lifetime. Generally, a larger production rate of the ALP  leads to a shorter lifetime, and vice versa. It is important to strike a balance between these two factors to effectively search for potential signals at either far or prompt detectors. 

\begin{figure*}[ht!]
\includegraphics[width=0.45\linewidth]{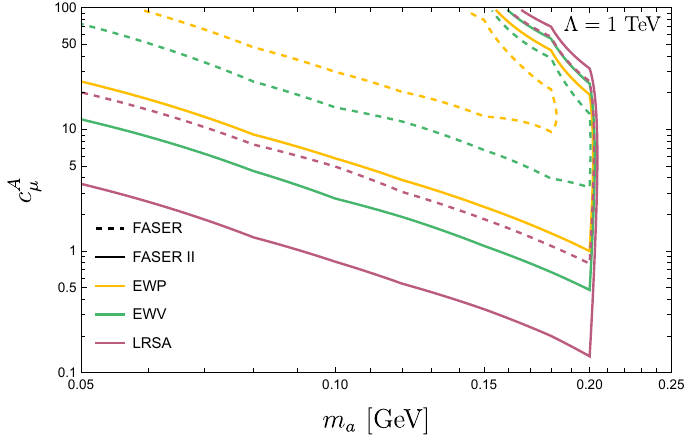}\qquad 
\includegraphics[width=0.45\linewidth]{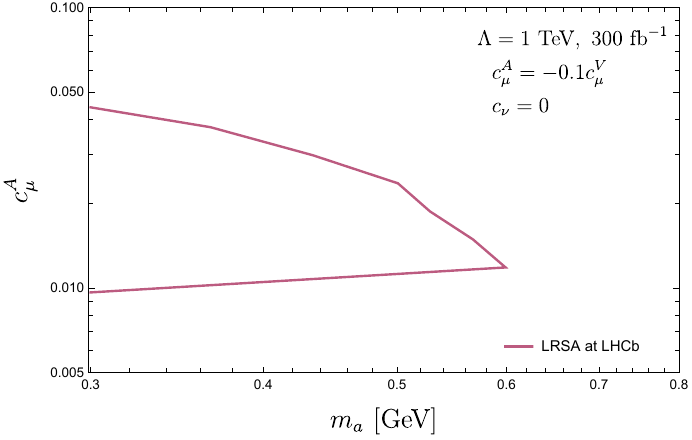} 
\caption{The $95\%$ confidence level projects for muonphilic ALPs. EWP, EWV and LRSA scenarios are presented in orange, green and violet lines, respectively. \textbf{Left panel:} FASER (dashed line) and FASER II (solid lines). \textbf{Right panel:} LHCb (solid line).}
\label{fig:muonfaserlhcb}
\end{figure*}

As discussed above, we consider the searches for muonphilic ALPs at both FASER and LHCb. The PDG data~\cite{ParticleDataGroup:2024cfk} are used as inputs. For the FASER analysis, we follow the same strategy introduced in Sec.~\ref{sec:es}, where Pythia 8, EPOS LHC, and the foresee package are employed to simulate ALP productions, decays, and detector effects. The branching ratios for hadron decays are taken from Sec.~\ref{sec:es}. Notably, within the ALP mass range of interest,  the 3-body decay of $\pi^\pm$ is kinematically forbidden. The remaining hadrons are considered inclusively, and we again impose the requirements that the total energy inside the detectors exceeds $500$ GeV. As before, we adopt the $N=3$ hypothesis to establish the $95\%$ confidence level for the background-free analysis. The exclusion limits on the $(m_a, c_\mu^A)$ plane are shown in the left panel of Fig.~\ref{fig:muonfaserlhcb}.

We find that all three scenarios, $i.e.$ EWP, EWV and LRSA, can be either detected or constrained at both FASER and FASER II. However, all $95\%$ confidence level contours terminate around $m_a \sim 2m_\mu$ consistent with our earlier discussions. Above this mass threshold, the dimuon decay is activated, significantly reducing the ALP lifetime. Notably, attempting to probe weaker coupling does not provide much benefit, as it simultaneously diminishes the significance by drastically lowering the ALP production rate.

The sensitivity projections of FASER and FASER II differ significantly from those of other searches. As noted in Ref.~\cite{Cheung:2022umw}, the search using supernova SN-1987A is restricted to the region where $m_a < 10$ MeV and $c_\mu^A \lesssim 20$. Super-Kamiokande imposes constraints requiring $c_\mu^A$ to exceed at least $100$ for $m_a < 30$ MeV, while BaBar is only effective when the ALP mass surpasses the $2m_\mu$ threshold. Taken together, FASER and FASER II are poised to explore largely uncharted regions of the $(m_a, c_\mu^A)$ parameter space, providing crucial complementarity to searches conducted at other experimental facilities.

For an ALP mass of $\mathcal O(0.1)$ GeV, the diphoton decay channel of a muonphilic ALP, rather than the dilepton decay, ensures a lifetime several orders of magnitude longer than that of its electrophilic counterpart, assuming LFU holds. In other words, for a given ALP lifetime, the diphoton decay allows a muonphilic ALP to achieve a significant larger coupling compared to an electrophilic ALP. This, in turn, ensures sufficient ALP production rates and event yields at detectors. As a result, even the EWP scenario can be effectively constrained at FASER, with even better projections achievable under scenarios such as EWV or LRSA.

For the LHCb analysis, we focus exclusively on the $B^0\to K^\ast a$ process to avoid overwhelming SM background events.
In this context, we only consider ALPs heavier than $2m_\mu$ and target the $a\to \mu^+\mu^-$ decay signals. Our simulation follows the methodology outlined in our previous work~\cite{Cheng:2024aco}, where the Hidden Valley module in Pythia 8 was utilized. The $B$ meson is required to reside in the forward region, satisfying $\eta \in [2,\ 5]$ and its production rate is normalized to the value measured by LHCb~\cite{LHCb:2017vec}, which aligns well with theoretical predictions~\cite{Mazzitelli:2023znt}. We assume that $B^0 (\bar B^0)$ and $B^\pm$ behave identically in this analysis without loss of generality. The recent LHCb dimuon search~\cite{LHCb:2020ysn}, based on $5.1$ fb$^{-1}$ data, is adopted as a benchmark. Specifically, the event selections for displaced vertex searches, as listed in Table 1 of Ref.~\cite{LHCb:2020ysn}, are applied at the truth level. Background events are extracted from the "inclusive" type (the search used in Ref.~\cite{LHCb:2020ysn}). The evaluation are done within $\pm 2\sigma$ range, where the yields are summed across eight $m_{\mu\mu}$ bins around $m_a$, as each bin width corresponds to half of the $m_{\mu\mu}$ resolution at LHCb. Compared to statistical uncertainties, the systematic uncertainties contribute barely $5\%$~\cite{LHCb:2020ysn} and therefore will be neglected in this work.

The high-luminosity phase, reaching $300$ fb$^{-1}$, is particularly promising. We believe its benefits extend beyond a simple luminosity rescaling, as significantly lower trigger thresholds and relatively reduced pile-up are expected starting from the LHC Run 3. Following the methodology outlined in Ref.~\cite{Ilten:2016tkc}, we adopt softer selection cuts, consistent with those applied in Ref.~\cite{Cheng:2024aco}. Specifically, we require both muon tracks to satisfy $p_T > 0.5$ GeV, $|p|> 10$ GeV, and $\eta \in [2,\ 5]$. Additionally, the reconstructed ALP must meet the criteria $p_T > 0.5$ GeV and transverse displacement $\ell_{xy} \in [6,\ 22]$ mm. To account for potential improvements in the $m_{\mu\mu}$ resolution, the mass window width is set to $16$ MeV for $m_{\mu\mu} < 1$ GeV, or $0.016 m_{\mu\mu}$ for $m_{\mu\mu}\geq 1$ GeV~\cite{Ilten:2016tkc}. Finally, a global detector efficiency of $0.7$ is universally applied as a benchmark, consistent with previous studies~\cite{Ilten:2016tkc, Cheng:2021kjg,Cheng:2024hvq, Cheng:2024aco}.

We summarize our findings from the LHCb analysis below. Unfortunately, even the HL-LHC is unable to project any constraints in the EWP and EWV scenarios. This aligns with our previous UV-independent study~\cite{Cheng:2024aco}. According to Fig.~15 in that study, for an ALP with $m_a \sim \mathcal O(10^2)$ MeV and $c\tau_a \sim \mathcal O(1)$ cm, the HL-LHC is effective only if the branching ratio BR$(B^0 \to K^\ast a)$ exceeds $10^{-10}$. In the muonphilic scenario, an ALP lifetime on the order of $\mathcal O(1)$ cm corresponds to a coupling $c_\mu^A \sim 0.01$, which results in a branching ratio approximately two orders of magnitude lower than the critical value, as derived from Fig.~\ref{fig:electrobraxiongeneration}~\footnote{Regarding muonphilic ALP production rates, a rescaling factor of $(\frac{m_\mu}{m_e})^2$ should be considered properly, as we've discussed in Sec.~\ref{sec:alpproduction}.}. However, opportunities emerge in the LRSA scenario, where ALP production is enhanced by a naturally non-vanishing $c_\mu^V$. The corresponding projection is shown in the right panel of Fig.~\ref{fig:muonfaserlhcb}. At the HL-LHC, the $95\%$ confidence level projection on the $(m_a, c_\mu^A)$ plane becomes achievable. The coupling $c_\mu^A$ can be constrained at $\mathcal O(0.01)$, while the ALP mass lies below $0.6$ GeV but above the threshold $2m_\mu$. Interestingly, the preferred parameter regions for LHCb differ significantly from those accessible to FASER and FASER II, highlighting complementarities in muonphilic ALP searches. Specifically, far detectors like FASER and FASER II primarily probe the diphoton decay, favoring parameter space with low masses (below $2m_\mu$) and relatively larger couplings. Conversely, prompt detectors like LHCb focus on dimuon decays, targeting regions with heavier masses and more feeble couplings, without any overlap with the far detector regions. 

\section{Conclusion}\label{sec:con}

This study investigates the detection prospects of leptophilic axion-like particles (ALPs) at forward detectors, focusing on ALPs with the electrophilic and muonphilic scenarios. By analyzing exotic meson decays with two-body and three-body final states involving ALPs, which subsequently decay into $e^+e^-$, $\mu^+\mu^-$, or $\gamma\gamma$, we provide a comprehensive assessment of the potential of detecting ALPs at FASER, FASER II, and LHCb.

For the electrophilic ALPs, we demonstrate that FASER II provides a significant improvement over FASER due to its increased detector size and integrated luminosity. This upgrade enables sensitivity to ALPs with masses up to $1$ GeV and couplings as small as $\mathcal{O}(1)$ which can be found in Fig.~\ref{fig:electrofaser}. The interplay between the dielectron and diphoton decay channels is critical for achieving these sensitivities, particularly in regions of parameter space that overlap with or complement existing constraints from experiments such as E137 (electron beam dump) and searches for flavor-changing neutral current (FCNC) decays at prompt detectors. Importantly, FASER II is poised to probe previously unexplored regions of parameter space, particularly for ALPs with low masses and weak couplings.

For the muonphilic ALPs, the higher ALP decay threshold, larger than  $2m_\mu$, and shorter ALP lifetimes present unique challenges. As shown in Fig.~\ref{fig:muonfaserlhcb}, far detectors such as FASER and FASER II remain sensitive to diphoton signals for lighter ALPs, while LHCb excels in probing heavier ALPs that decay promptly into dimuons. Our analysis indicates that LHCb’s high-luminosity phase will allow for robust constraints in the left-right softly asymmetric (LRSA) scenario, where non-vanishing vector couplings naturally enhance ALP production rates. The complementarity between far detectors and prompt detectors is particularly striking: while far detectors excel in searching for long-lived ALPs with small masses and large couplings, prompt detectors like LHCb are better suited for heavier ALPs with feeble couplings.

\acknowledgments{We thank Kai-Feng Zheng, Ke Li, Zeren Simon Wang and Wei Liu for useful discussions. X.J. and C.L. were supported in part by the National Natural Science Foundation of China (NNSFC) under grant No.~12342502 and No.~12335005, respectively. C.L. was also supported by the Special funds for postdoctoral overseas recruitment, Ministry of Education of China.
}

\appendix

\section{ALP-Lepton Interactions}\label{app: interaction}

In this appendix, we will focus on the derivation of leptophilic ALP interactions, following the convention introduced in Refs.~\cite{Altmannshofer:2022izm, Lu:2022zbe}. Generally, by introducing a cut-off scale $\Lambda$, the Lagrangian takes the form $\mathcal L = \partial_\mu a j^\mu_{PQ}$, with the Peccei-Quinn(PQ) current as:
\begin{equation}
    j^\mu_{PQ} = \frac{c_\ell^V}{2\Lambda} \bar\ell \gamma^\mu \ell + \frac{c_\ell^A}{2\Lambda} \bar\ell \gamma^\mu \gamma^5 \ell + \frac{c_\nu}{2\Lambda} \bar\nu_\ell \gamma^\mu P_L \nu_\ell ~,
\end{equation}
where $P_L$ represents the chiral projection operator, defined as $P_{R(L)}=\frac{1}{2}(1\pm \gamma^5)$.

To reformulate the interactions between the ALP and the leptonic sector, integration by parts is applied to shift the derivative to the PQ current, expressed as $\partial_\mu j^\mu_{PQ}$. The three components, namely the vector, pseudo-vector, and neutrino currents, contribute differently and are analyzed individually.

First and foremost, the vector current can be reformulated into two separate terms based on lepton chirality, as shown in the following expression: 
\begin{align}\label{eq: derivative}
\partial_\mu j^\mu_{PQ} & \supset \frac{c_\ell^V}{2\Lambda} \partial_\mu (\bar\ell_L \gamma^\mu \ell_L + \bar\ell_R \gamma^\mu \ell_R) \\ \nonumber
&= \frac{c_\ell^V}{2\Lambda} \bigg [ \bar\ell_L \gamma^\mu (\partial_\mu \ell_L) + (\partial_\mu \bar\ell_L) \gamma^\mu \ell_L \bigg ] + \frac{c_\ell^V}{2\Lambda}\partial_\mu (\bar\ell_R \gamma^\mu \ell_R) \\ \nonumber
& = \frac{c_\ell^V}{2\Lambda}\bar\ell_L \gamma^\mu \bigg [(D_\mu + i G_\mu^C + i G_\mu^N) Q_L \bigg ]_{\rm down} + ... \\ \nonumber
& = \frac{ig c_\ell^V}{2\sqrt{2} \Lambda} (\bar \ell_L \gamma^\mu \nu_{\ell L} W_\mu^-) + h.c..
\end{align}
The first line displays the decomposition of the current, while the third line reinterprets it by incorporating the SM $SU(2)_L$ gauge structure. Here, "down" refers to taking the lower component of the $SU(2)_L$ doublet, and the ellipsis denotes additional terms that can be treated similarly. $D_\mu$ represents the gauge covariant derivative for left-chiral fermions, while $G_\mu^C$ and $G_\mu^N$ correspond to contributions from charged and neutral gauge bosons, respectively. These terms are defined as:
\begin{align}
D_\mu &= \partial_\mu - i G_\mu^C - i G_\mu^N ~, \\
G_\mu^C &= \frac{g}{\sqrt{2}}(W_\mu^+T^+ + W_\mu^-T^-) ~, \\
G_\mu^N &= \frac{g}{c_W}Z_\mu (T^3 - s_W Q_e) + e Q_e A_\mu~,
\end{align}
where $W_\mu^\pm$, $Z_\mu$ and $A_\mu$ are the gauge bosons, $g$ is the $SU(2)_L$ gauge coupling, $c_W$ and $s_W$ represent the cosine and sine of the Weinberg angle, and $eQ_e$ denotes the electric charge. In addition, the matrices $T^3$ and $T^\pm$ are defined as follows: 
\begin{equation*}
T^3 = \big (\begin{smallmatrix}
  1 & 0\\ 
  0 & -1
\end{smallmatrix}\big )~,~T^+ = \big (\begin{smallmatrix}
  0 & 1\\ 
  0 & 0
\end{smallmatrix}\big )~,~T^- = \big (\begin{smallmatrix}
  0 & 0\\ 
  1 & 0
\end{smallmatrix}\big )~.
\end{equation*}

The equation of motion (EOM),
\begin{equation}\label{eq: rom}
i D_\mu \gamma^\mu Q_L = \Gamma_\ell \ell_R H ~,
\end{equation}
is then applied, where $\Gamma_\ell$ is the Yukawa matrix for the lepton sector and $H$ stands for the Higgs doublet. To neglect interactions with Higgs boson, only the vacuum expectation value is considered, $i.e.$ $H = (0,v)^T$. The right-chiral part is handled in a very similar way, though it is much simpler due to the absence of $W^\pm$ boson interactions. After careful calculations, the final result appears in the fourth line of Eq.~(\ref{eq: derivative}). In particular, all terms involving the photon or $Z$ boson cancel out, leaving only the $W^\pm$ bosons relevant to this study.

Secondly, the analysis of the pseudo-vector current follows a similar procedure, with the main difference being the replacement of the combination $P_{R(L)}$ from $P_R + P_L$ to $P_R - P_L$. The resulting expression is: 
\begin{align}
    \partial_\mu j^\mu_{PQ} &\supset i c_\ell^A \frac{m_\ell}{\Lambda} \bar \ell \gamma^5 \ell \nonumber \\
    &-\frac{ig c_\ell^A}{2\sqrt{2} \Lambda} (\bar \ell_L \gamma^\mu \nu_{\ell L} W_\mu^-) + h.c. ~.
\end{align}
Finally, the treatment of the neutrino current is even simpler. In this case, only the upper component of the $SU(2)_L$ doublet is relevant, and the right-hand side of the neutrino EOM vanishes. Therefore, the result is:
\begin{equation}
    \partial_\mu j^\mu_{PQ} \supset -\frac{ig c_\nu}{2\sqrt{2} \Lambda} (\bar \ell_L \gamma^\mu \nu_{\ell L} W_\mu^-) + h.c. ~.
\end{equation}

Furthermore, ALPs can couple to gauge bosons when the chiral anomaly effect is taken into account. The photon term, associated with the Adler-Bell-Jackiw anomaly~\cite{Adler:1969gk,Bell:1969ts}, is given by:
\begin{equation}
\partial_\mu j^\mu_{PQ,~AA} = -\frac{c_\ell^A e^2}{16\pi^2\Lambda} F_{\mu\nu} \tilde F^{\mu\nu} ~,
\end{equation}
where $\tilde F^{\mu\nu} \equiv \frac{1}{2} \epsilon^{\alpha\beta\mu\nu} F_{\alpha\beta}$ and $\epsilon^{0123}=1$. It is worth noting that the vector current remains conserved at the both the classical and quantum levels, contributing nothing to the derivative. In contrast, the pseudo-vector current becomes relevant due to the Feynman diagram involving a triangle fermionic loop.

The terms involving other gauge bosons can be derived analogously to the photon case, starting with the mathematical structure of the $\gamma$-matrices. In the photon case, the loop diagram induced by the PQ current takes the following form:
\begin{equation}\label{eq:structure}
\partial_\mu j^{\mu}_{PQ,~AA} \sim \int d^4 p~\textbf{Tr}\bigg [p_\mu \gamma^\mu \gamma^5 (\slash{p}+m_\ell) \gamma^\nu (\slash{k}+m_\ell) \gamma^\rho (\slash{q}+m_\ell)\bigg ] ~,
\end{equation}
where $p$, $k$ and $q$ are the lepton momenta in the loop. To keep the expression concise, we do not explicitly include the denominator here. A key point to note is that the trace of $\gamma^5$ 
multiplied by an odd number of $\gamma^\mu$ matrices vanishes. Therefore, in Eq.~(\ref{eq:structure}), only terms with six $\gamma^\mu$-matrices and four $\gamma^\mu$-matrices contribute. However, for the latter case (with four $\gamma^\mu$), the integral becomes negligible when comparing the powers of $p$ in the numerator and denominator. As a result, we can discard the trivial contributions and simplify the expression to: 
\begin{align}
\partial_\mu j^{\mu}_{PQ,~AA} \sim \int d^4 p~\textbf{Tr}\bigg [p_\mu \gamma^\mu \gamma^5 \slash{p} \gamma^\nu \slash{k}\gamma^\rho \slash{q}\bigg ] ~,
\end{align}
without loss of generality.

Meanwhile, some straightforward calculations yield the following relations:  
\begin{align}
& \textbf{Tr}\bigg [...\slash{p}~(\gamma^\nu P_{L(R)})~\slash{k} ~(\gamma^\rho P_{L(R)}) ~\slash{q} \bigg ] \notag \\
&= \textbf{Tr}\bigg [...\slash{p}\gamma^\nu P_{L(R)} \slash{k} \gamma^\rho \slash{q}\bigg ]~,
\end{align}

\begin{align}
& \int d^4p~ \textbf{Tr}\bigg [p_\mu \gamma^\mu \slash{p}\gamma^\nu P_{L(R)}\slash{k} \gamma^\rho \slash{q}\bigg ] \notag \\
&= \mp \frac{1}{2}\int d^4p~ \textbf{Tr}\bigg [p_\mu \gamma^\mu \gamma^5 \slash{p}\gamma^\nu \slash{k} \gamma^\rho \slash{q}\bigg ]~, \nonumber \\
& \int d^4p~ \textbf{Tr}\bigg [p_\mu \gamma^\mu \gamma^5 \slash{p}\gamma^\nu P_{L(R)}\slash{k} \gamma^\rho \slash{q}\bigg ] \notag \\
&= \frac{1}{2}\int d^4p~ \textbf{Tr}\bigg [p_\mu \gamma^\mu \gamma^5 \slash{p}\gamma^\nu \slash{k} \gamma^\rho \slash{q}\bigg ]~.
\label{eq:anal1}
\end{align}
In Eq.~(\ref{eq:anal1}), we have applied the fact that the vector current contribution in Eq.~(\ref{eq:structure}) vanishes. Using these identities, we can proceed analogously by considering the chiral properties and replacing the QED vertex ($-ie\gamma^\mu$) with its weak interaction counterparts. For instance, for the vertex involving left-chiral leptons and a $Z$ boson, we have:
\begin{equation}
   -ie\gamma^\mu \rightarrow -\frac{ie}{s_W c_W}(\frac{1}{2}-s_W^2)\gamma^\mu P_L~,
\end{equation}
where the relation $e\equiv g s_W$ has been adopted.

The same strategy can be applied to terms involving $Z_{\mu\nu}\tilde Z^{\mu\nu}$, $F_{\mu\nu}\tilde Z^{\mu\nu}$, and, $W_{\mu\nu}^+\tilde W^{-,\mu\nu}$. In particular, for the case $F_{\mu\nu}\tilde Z^{\mu\nu}$, since the photon and $Z$ boson are not identical, an additional factor of $2$ must be included. As a result, we obtain:
\begin{align}
\partial_\mu j^\mu_{PQ} & \supset \frac{e^2}{16\pi^2\Lambda}\bigg [ \frac{c_\ell^V - c_\ell^A + c_\nu}{4s_W^2} W_{\mu\nu}^+\tilde W^{-,\mu\nu} \nonumber \\
& + \frac{c_\ell^V - c_\ell^A(1-4 s_W^2)}{2c_W s_W}F_{\mu\nu}\tilde Z^{\mu\nu} - c_\ell^A F_{\mu\nu}\tilde F^{\mu\nu} \nonumber \\
& + \frac{c_\ell^V(1-4 s_W^2)-c_\ell^A(1-4 s_W^2+8s_W^4)+c_\nu}{8s_W^4c_W^4}Z_{\mu\nu}\tilde Z^{\mu\nu}\bigg ]~.
\end{align}

Combining all the results, we obtain:
\begin{align}
a\cdot \partial_\mu j^\mu_{PQ} &= i c_\ell^A \frac{m_\ell}{\Lambda} a\bar \ell \gamma^5 \ell \nonumber \\
& + \frac{e^2}{16\pi^2\Lambda}\bigg [ \frac{c_\ell^V - c_\ell^A + c_\nu}{4s_W^2} aW_{\mu\nu}^+\tilde W^{-,\mu\nu} \nonumber \\
& + \frac{c_\ell^V - c_\ell^A(1-4 s_W^2)}{2c_W s_W}aF_{\mu\nu}\tilde Z^{\mu\nu} - c_\ell^A aF_{\mu\nu}\tilde F^{\mu\nu} \nonumber \\
& + \frac{c_\ell^V(1-4 s_W^2)-c_\ell^A(1-4 s_W^2+8s_W^4)+c_\nu}{8s_W^4c_W^4}aZ_{\mu\nu}\tilde Z^{\mu\nu}\bigg ] \nonumber \\
& - ig\frac{c_\ell^A - c_\ell^V + c_\nu}{2\sqrt{2} \Lambda} (\bar \ell_L \gamma^\mu \nu_{\ell L} W_\mu^-a) + h.c. ~. 
\end{align}
The first line describes a three-point interaction involving the ALP and leptons. The second to fourth lines correspond to interactions with gauge bosons, which arise due to chiral anomalies.  The final line introduces a novel four-point interaction involving a charged lepton, a neutrino, a $W$ boson and an ALP.


\end{document}